\documentclass[prb,twocolumn,floatfix,showpacs,amsmath,amssymb]{revtex4}

\usepackage{graphicx,color}
\usepackage{dcolumn}
\usepackage{bm}
\usepackage[latin1]{inputenc}

\newcommand{\lao}{LaFeAsO}
\newcommand{\laf}{LaFeAsO$_{1-x}$F$_x$}
\newcommand{\lafuenf}{LaFeAsO$_{0.95}$F$_{0.05}$}
\newcommand{\lazehn}{LaFeAsO$_{0.9}$F$_{0.1}$}
\newcommand{\rf}{REFeAsO$_{1-x}$F$_x$}
\newcommand{\tc}{$T_{\rm{C}}$}
\newcommand{\tn}{$T_{\rm{N}}$}

\newcommand{\tsdw}{$T_{\rm{N}}$}
\newcommand{\ts}{$T_{\rm{S}}$}

\newcommand{\figref}[1]{Fig.\,\protect\ref{#1}}

\newcommand{\cax}{Ca(Fe$_{1-x}$Co$_x$)$_2$As$_2$}

\begin{document}

\title{Local antiferromagnetic correlations in the iron pnictide superconductors LaFeAsO$_{1-x}$F$_x$ and Ca(Fe,Co)$_2$As$_2$
as seen via normal-state susceptibility}

\author{R.~Klingeler}\affiliation{Institute for Solid State Research, IFW Dresden, D-01171 Dresden, Germany}
\author{N.~Leps}\affiliation{Institute for Solid State Research, IFW Dresden, D-01171 Dresden, Germany}
\author{I.~Hellmann}\affiliation{Institute for Solid State Research, IFW Dresden, D-01171 Dresden, Germany}
\author{A.~Popa}\affiliation{Institute for Solid State Research, IFW Dresden, D-01171 Dresden, Germany}
\author{U.~Stockert}\affiliation{Institute for Solid State Research, IFW Dresden, D-01171 Dresden, Germany}
\author{C.~Hess}\affiliation{Institute for Solid State Research, IFW Dresden, D-01171 Dresden, Germany}
\author{V.~Kataev}\affiliation{Institute for Solid State Research, IFW Dresden, D-01171 Dresden, Germany}
\author{H.-J.~Grafe}\affiliation{Institute for Solid State Research, IFW Dresden, D-01171 Dresden, Germany}
\author{F.~Hammerath}\affiliation{Institute for Solid State Research, IFW Dresden, D-01171 Dresden, Germany}
\author{G.~Lang}\affiliation{Institute for Solid State Research, IFW Dresden, D-01171 Dresden, Germany}
\author{G.~Behr}\affiliation{Institute for Solid State Research, IFW Dresden, D-01171 Dresden, Germany}
\author{L.~Harnagea}\affiliation{Institute for Solid State Research, IFW Dresden, D-01171 Dresden, Germany}
\author{S.~Singh}\affiliation{Institute for Solid State Research, IFW Dresden, D-01171 Dresden, Germany}
\author{B.~B\"uchner}\affiliation{Institute for Solid State Research, IFW Dresden, D-01171 Dresden, Germany}


\date{\today}

\begin{abstract}
We have studied the interplay of magnetism and superconductivity in LaFeAsO$_{1-x}$F$_x$ and \cax
. While antiferromagnetic spin density wave formation is suppressed and superconductivity evolves,
all samples show a doping independent strong increase of the normal state susceptibility upon
heating which appears a general feature of iron pnictides. The data provide evidence for robust
local antiferromagnetic correlations persisting even in the superconducting regime of the phase
diagram.
\end{abstract}



\maketitle



The appearance of high-temperature superconductivity (SC) in \laf\ and related iron pnictides \rf\
(with RE = Ce,Nd,Sm,Gd) and AFe$_{2}$As$_{2}$ (A = Ca,Sr,Ba) renews strong interest in the complex
interplay of magnetism and
SC.~\cite{Kamihara2008,Chen2008,Chen2008a,Ren2008,Rotter2008b,Sasmal2008} Similar to cuprate
high-\tc\ superconductors, iron pnictides exhibit SC in magnetic layers of 3$d$-atoms, i.e.
Fe-atoms, which form a regular square lattice. In the parent material \lao , antiferromagnetic
(afm) spin density wave (SDW) order evolves below \tn\ $\sim$ 140~K and a structural transition
from tetragonal to orthorhombic symmetry occurs at \ts $\sim$
160~K.~\cite{Nomura2008,Cruz2008,Dong2008,Klauss2008} Like in the cuprates, SC is associated with
suppression of magnetic order by electron or hole doping but the undoped parent material is not a
Mott-Hubbard insulator but a multi-band metal. In superconducting \laf , i.e. for $x\geq 0.05$,
the presence of magnetic order has been ruled out experimentally.~\cite{Luetkens2008,Luetkens2009}
Introducing additional magnetic moments can induce slow spin fluctuations and even static
magnetism in the vicinity of the superconducting phase of SmFeAsO$_{1-x}$F$_x$ \cite{Drew} but
such fluctuations are clearly absent in the La-based compounds. NMR data show that the spin
dynamics in \laf\ varies markedly with F doping since afm fluctuations which are clearly visible
at $x=0.04$ disappear for $x=0.11$.~\cite{Nakai2008} In contrast, our data show that local afm
correlations can be present in superconducting \laf\ and \cax\ up to the overdoped regime. This
conclusion is drawn from the temperature dependence of the normal state susceptibility $\chi_{\rm
norm}$ in \laf\ with $x\geq 0.05$ as well as in superconducting \cax\ single crystals. Despite the
suppression of the afm order and the occurrence of SC, $\chi_{\rm norm}$ is still increasing upon
heating and exhibits the same slope as in the undoped case. The robustness and generic character
of this behavior at elevated temperatures is remarkable since the ground state qualitatively
changes upon doping from a magnetically ordered state to a superconducting state.


Polycrystalline samples of $\rm LaO_{1-x}F_xFeAs$ with $0\leq x\leq 0.15$ were prepared from pure
components as described in Ref.~\onlinecite{Kondrat}. Single crystals of \cax\ ($0\leq x\leq
0.125$) have been grown in Sn flux as will be described in detail elsewhere.~\cite{Surjeet} The
crystal structure and the composition were investigated by powder X-ray diffraction and
wavelength-dispersive X-ray spectroscopy (WDX). From the XRD data impurity concentrations smaller
than 3\% are inferred. In addition, our samples have been characterised in detail by, e.g.,
magnetisation, electrical resistivity, NMR, microwave and $\mu$SR
studies.~\cite{Luetkens2008,Luetkens2009,Narduzzo2008,Grafe2008,Hess_PhD,Klauss2008} Magnetization
measurements have been performed in a MPMS-XL SQUID magnetometer (Quantum Design) in the
temperature range 2--350\,K in a magnetic field of 1\,T and between 2\,K and 50\,K in 2\,mT.
Superconductivity was found for \laf\ with $x\geq 0.05$ and \cax\ with $x\geq 0.045$. For some
compositions, a tiny ferromagnetic contribution was detected which indicates a magnetic impurity
phase up to $\sim 0.01$\%. In the temperature range under study, the corresponding moment is
temperature independent and was subtracted from the susceptibility data. For \lao , measurements
up to 800\,K imply a temperature hysteresis of the magnetisation above $\sim$500\,K which we
attribute to a starting degradation of the sample. Upon F-doping, the onset of this effect is
found already at lower temperatures, i.e. down to 330\,K. For NMR experiments, oriented powder of
\lazehn\ was formed by grinding the material, mixing with Stycast 1266 epoxy and curing in an
external field of 9.2 T~\cite{Grafe2008}, while for \lafuenf\ a powder sample was studied. The NMR
measurements have been done in a fixed magnetic field of 7.0494 T.


\begin{figure}[htb]
\includegraphics [angle=0,width=0.95\columnwidth,clip] {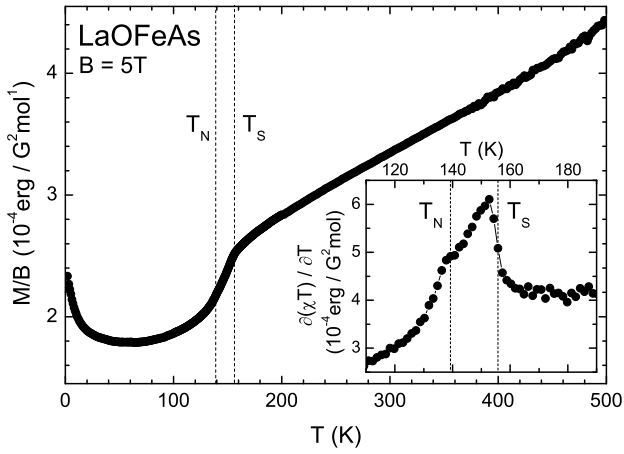}
\caption{Static susceptibility $\chi=M/B$ vs. temperature of \lao\ at $B=5$\,T. \ts\ and \tn\
label the structural and the afm phase transition, respectively. Inset: Magnetic specific heat
$\partial(\chi T)/\partial T$ in the vicinity of \ts .} \label{fig1}
\end{figure}

The temperature dependence of the static susceptibility $\chi=M/B$ in \lao\ (\figref{fig1}) shows
clear anomalies around $\sim$150\,K, similar to data in
Refs.~[\onlinecite{McGuire2008,Nomura2008,Kohama2008a}]. These anomalies are associated with the
structural phase transition to orthorhombic symmetry and to the onset of long range afm ordering,
respectively.~\cite{Klauss2008} Both phase transitions are clearly visible in the magnetic
specific heat which is proportional to $\partial(\chi T)/\partial T$ (inset of \figref{fig1}). The
anomaly at \ts\ = 156\,K indicates an enhancement of afm correlations at the structural phase
transition which demonstrates an intimate coupling between structure and magnetism. The anomaly of
the magnetic specific heat at \tn\ = 138\,K is qualitatively very similar, but much weaker than
the one at \ts . Remarkably, there is a pronounced increase of the susceptibility at higher
temperatures, i.e. the susceptibility increases upon heating up to at least 500\,K. This strong
temperature dependence by nearly a factor of 2 excludes conventional Pauli- and Curie-Weiss-like
paramagnetic behavior.

\begin{figure}
\includegraphics [width=0.95\columnwidth,clip] {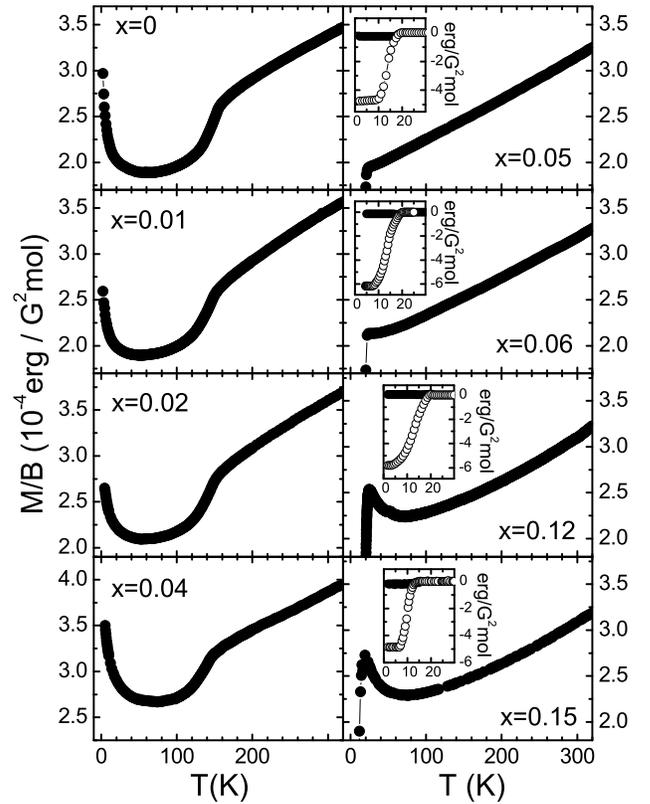}
\caption{Static susceptibility $\chi=M/B$ of \laf , for different doping level between 0$\leq x
\leq$ 0.15 at $B=1$\,T. Note, that for all graphs the ordinate covers the range $\Delta\chi =
2\cdot 10^{-4}$\,erg/G$^2$mol. Insets: $M$ vs $T$ for $B=2$\,mT.} \label{fig2}
\end{figure}

The effect of doping on the magnetic properties of \laf\ is presented in \figref{fig2}. For low
doping levels $x\leq 0.04$ there is only a moderate effect of F-doping on the structural and the
magnetic transition. In this low doping regime, the anomalies at the structural and magnetic phase
transitions as well as the linear increase of $\chi$ for $T
> T_{\rm S}$ are still visible. The susceptibility curves drastically change when
the doping level $x=0.05$ is achieved where a transition to the superconducting phase is observed
at \tc\ $\sim 19$\,K. In all superconducting samples there is no signature of the anomalies
associated to the structural and the magnetic phase transitions. Indeed, the presence of any
static magnetic order or structural effects in both underdoped and optimally doped samples was
ruled out by recent $\mu$SR and thermal expansion studies.~\cite{Luetkens2009,Wang2009} In
addition, $\chi(T)$ shows no signature of a Curie contribution due to paramagnetic impurity spins,
either intrinsic or belonging to a spurious phase. In stark contrast to the strong effect on the
ordered phase, the doping effect on the paramagnetic susceptibility is however negligible. For
$x=0.05$, e.g., there is a nearly perfect linear $\chi(T)$ dependence in the temperature regime
between $\sim$30\,K up to at least 380\,K. For higher F-concentrations, this behavior is
superimposed by a susceptibility upturn in the vicinity of $T_{\rm C}$ but there are only
negligible changes at room temperature.

\begin{figure}
\includegraphics [width=0.85\columnwidth,clip] {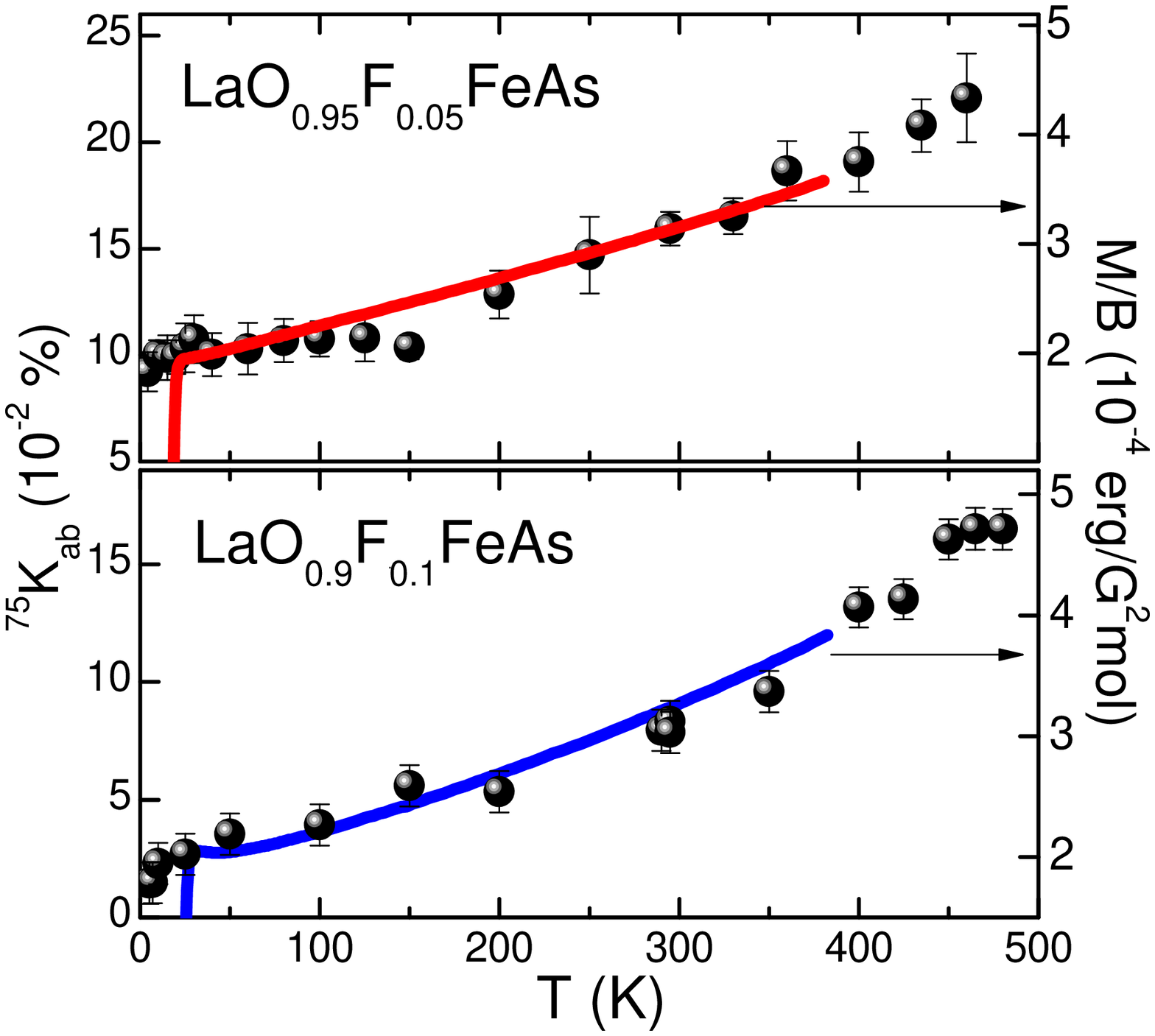}
\caption{(Color online) Knight shift $K_{ab}$ of $^{75}$As (circles) and the macroscopic
susceptibility of \lafuenf\ and \lazehn\ versus temperature with different vertical scales and
origins.} \label{knight}
\end{figure}

In order to prove the intrinsic electronic origin of the experimentally observed static
magnetization data we have investigated the electronic spin susceptibility of the FeAs layers in
\laf , with $x=0.05$ and $x=0.1$, as probed by the $^{75}$As nuclear magnetic resonance (NMR)
(\figref{knight}). Similar to the $\chi(T)$-dependence, the in-plane magnetic shift of the
$^{75}$As NMR signal $K_{ab}$ linearly increases from \tc\ up to room temperature. We observe a
clear scaling of $K_{ab}$ and macroscopic $\chi_{\rm norm}$ as shown in \figref{knight}, where the
ordinate scales are adjusted to match the curves in the paramagnetic regime. In general, the
magnetic shift $K_{ab}$ consists of the Knight (spin) shift $K_s \propto \chi_{\rm spin}$ due to
the hyperfine coupling to the electron spins and the orbital shift $K_{orb}$ which reflects a
$T$-independent orbital contribution. The fact that the NMR shift scales with the static
susceptibility above \tc\ implies that the observed macroscopic $\chi(T)$ is determined by the
intrinsic spin susceptibility $\chi_{\rm spin}$. Furthermore, similar to $K_s$ and $\chi$ the
$^{75}$As longitudinal relaxation rate $1/T_1$ divided by temperature shows, for $x=0.1$, an
increase~\cite{Grafe2008}, roughly yielding a typical for metals constant Korringa product $S =
K^2_sT_1T$\,=\,const \cite{Korringa}. This suggests that $^{75}$As nuclear moments are coupled to
itinerant quasiparticles and that itinerant electrons give the main contribution to the static
$\chi$.

\begin{figure}
\includegraphics [width=0.95\columnwidth,clip] {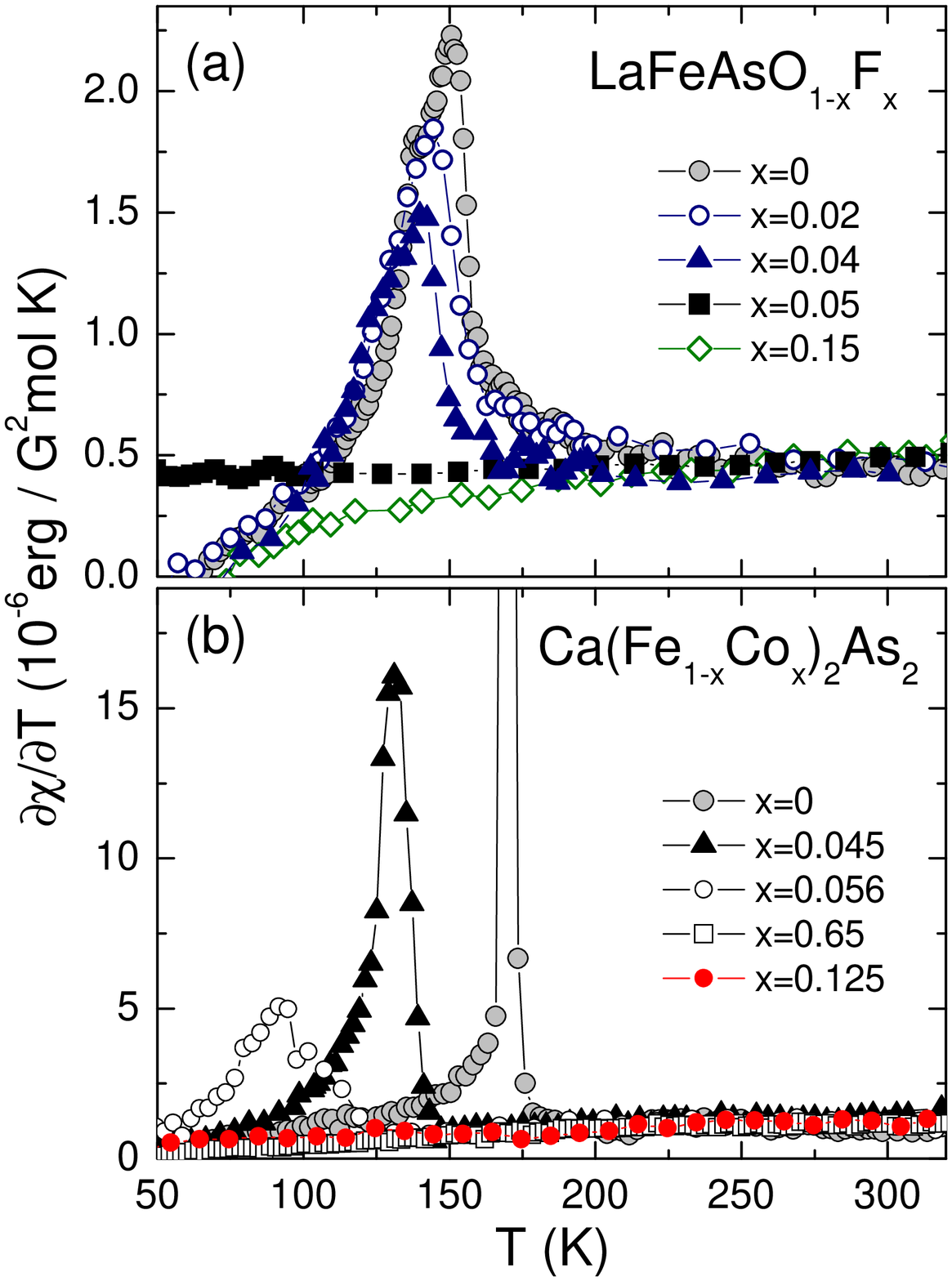}
\caption{Derivative of the static susceptibility $\partial \chi /\partial T$, at $B=1$\,T, of
\laf\ (a) and \cax\ (b), for selected doping levels.} \label{deriv}
\end{figure}

\begin{figure}
\includegraphics [width=0.85\columnwidth,clip] {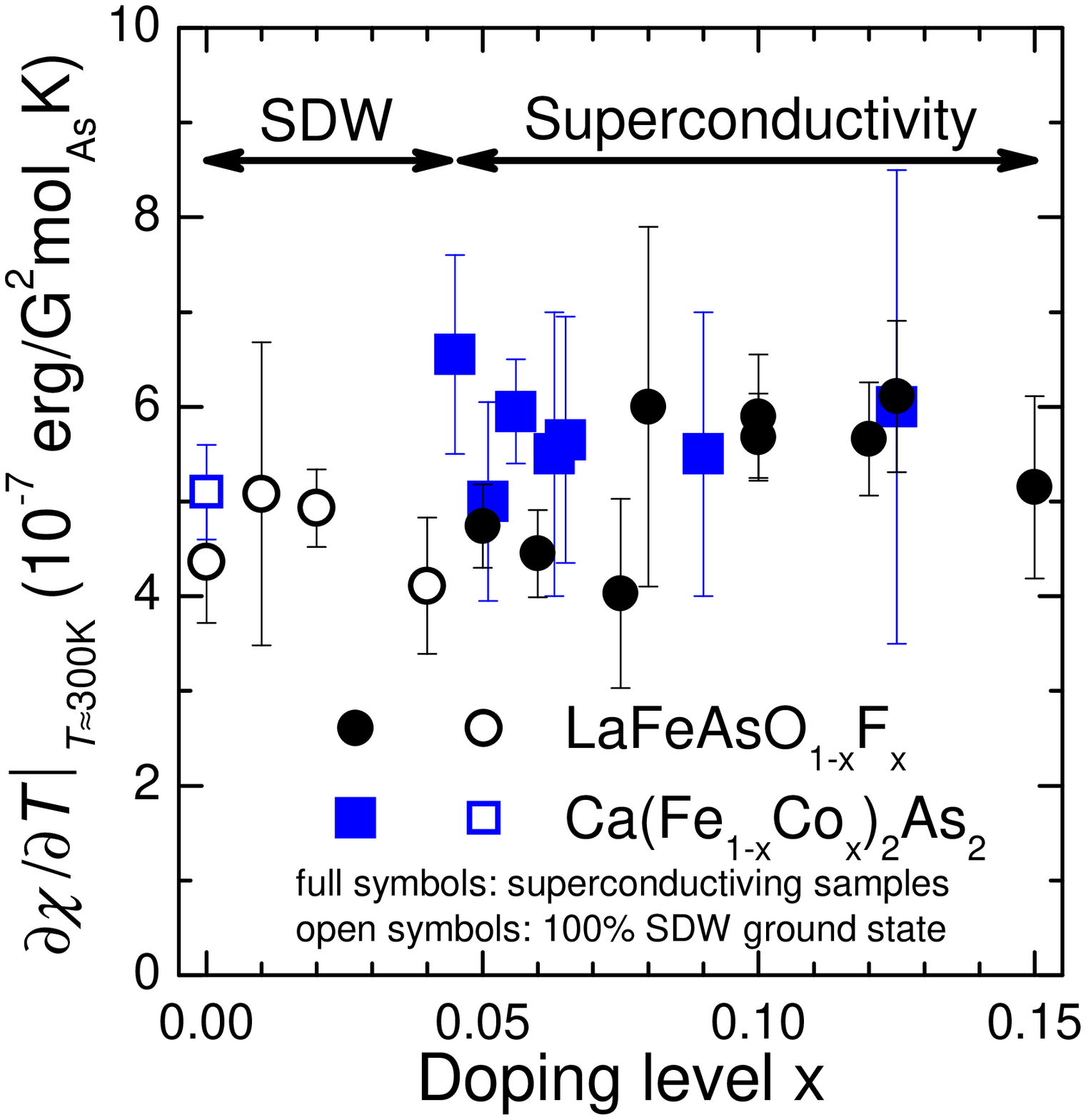}
\caption{(Color online) Slope of the susceptibility $\partial \chi /\partial T$, at 300\,K, as a
function of doping in \laf\ (circles) and \cax\ (squares). Filled symbols correspond to
superconducting samples.} \label{fig3}
\end{figure}

The main result of our susceptibility study is the observation of a similar slope of $\chi(T)$ in
the normal state for all doping levels under study. This behavior is already evident from the data
shown in \figref{fig2}. A quantitative analysis of the doping effect on the temperature dependence
around room temperature is presented in \figref{fig3}(a), where $\partial \chi /\partial T$ is
shown. For low doping levels $x\leq 0.4$, the data exhibit large anomalies which indicate the
structural and the SDW transition, respectively. At high temperatures, the data imply $\partial
\chi /\partial T \approx 5\cdot 10^{-7}$\,erg/(G$^2$mol$\cdot$K). Note, that in this low doping
regime there are additional contributions to $\partial \chi /\partial T$ above \ts , e.g., up to
$\gtrsim 200$\,K for \lao , which are associated with concomitant structural and electronic
fluctuations.~\cite{Wang2009} Upon doping, the onset temperatures of these fluctuations as well as
\ts\ and \tn\ are suppressed. In agreement with the complete absence of the orthorhombic and
magnetic phases, there are no anomalies in $\partial \chi /\partial T$ for $x=0.05$, i.e.
$\chi(T)\propto T$. Here, the slope of $\chi_{\rm norm}(T)$ agrees to the one in the tetragonal
phase of the low doped compounds. For higher F-concentrations, the additional susceptibility
upturn in the vicinity of $T_{\rm C}$ yields a temperature regime with $\partial \chi /\partial T<
5\cdot 10^{-7}$\,erg/(G$^2$mol$\cdot$K), which onset increases upon doping from $\sim 40$\,K at
$x=0.06$ to $\sim 175$\,K for the highest investigated doping level $x=0.15$. At higher
temperatures, however, \textit{all} doping levels 0 $\leq x \leq$ 0.15 exhibit the same slope of
around $5\cdot 10^{-7}$\,erg/(G$^2$mol$\cdot$K). This is highlighted by the data in \figref{fig3},
which shows the slope of the susceptibility $\partial \chi /\partial T$, at 300\,K, as a function
of doping. Within the error bars, we there is no significant dependence on the doping level $x$.

\begin{figure}
\includegraphics [width=0.95\columnwidth,clip] {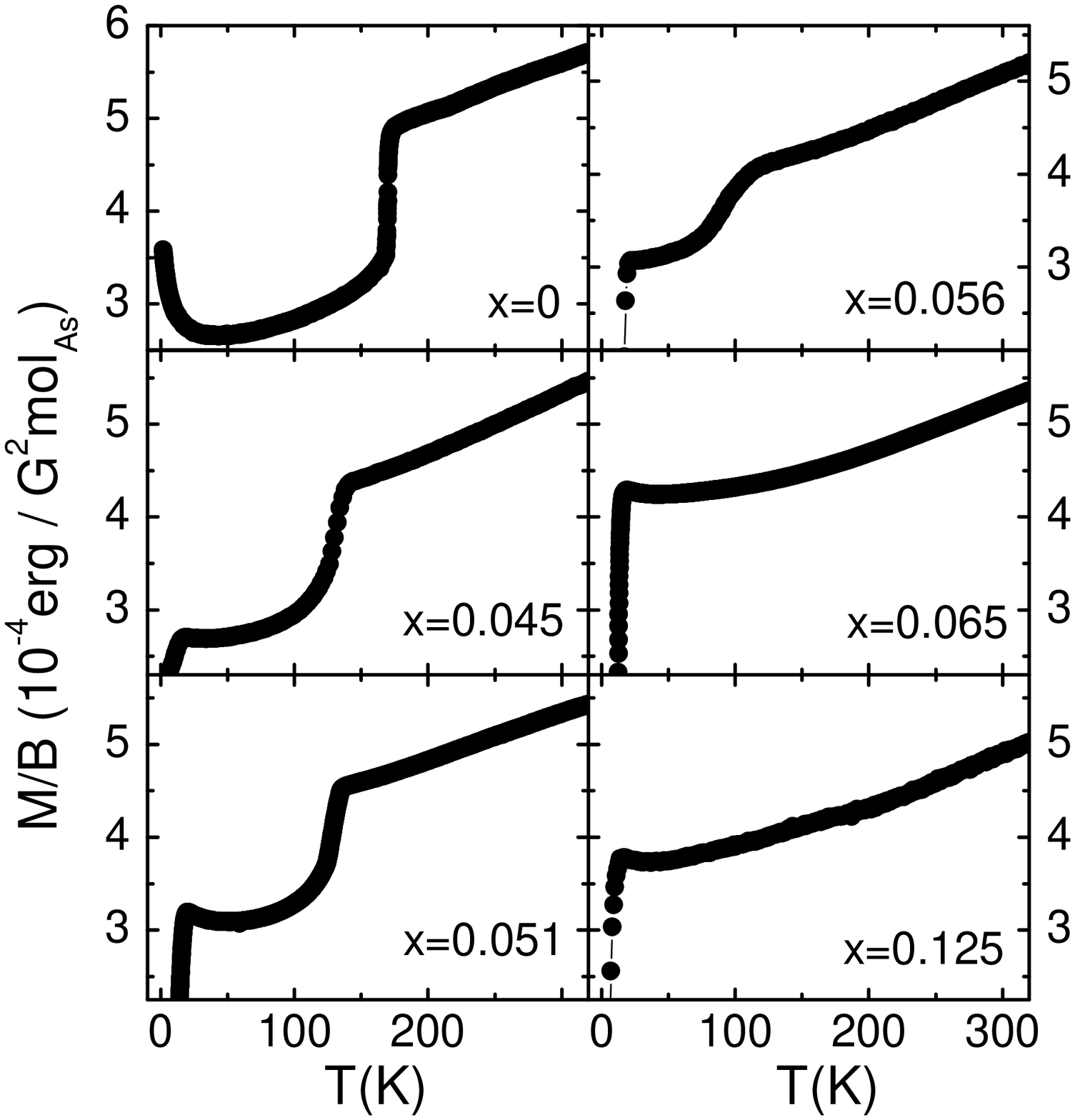}
\caption{Static susceptibility $\chi=M/B$ of \cax , for $0\leq x\leq 0.125$, at $B\|ab=1$\,T. For
all graphs the ordinate covers the range $\Delta\chi = 3.5\cdot 10^{-4}$\,erg/G$^2$mol.}
\label{xtals}
\end{figure}

Linear temperature dependence of the normal state susceptibility at elevated temperatures seems to
be a general feature of iron pnictides. This has been shown, e.g., for undoped 122-materials
AFe$_{2}$As$_{2}$
(A=Ca,Sr,Ba).~\cite{WangPRL2009,Zhang2009EPL,YanPRB2008,Ronning2008,Kumar2009PRB} As in the case
of \laf , however, this feature is not restricted to the undoped materials but persists in the
superconducting regime of the phase diagram as displayed by the susceptibility data on \cax\
($0\leq x\leq 0.125$) in \figref{xtals}. For $x=0$, linear behavior is found down to \tsdw\ =
171\,K, where our crystal exhibits a first-order transition to the SDW-state, which is in a good
agreement to Refs.\,[\onlinecite{Ronning2008,Kumar2009PRB}]. Upon Co-doping, the crystals become
superconducting. In addition, anomalies well above \tc\ indicate the structural and magnetic phase
transitions which temperature at different doping levels is also visible in $\partial \chi
/\partial T$ (\figref{deriv}(b)). Albeit this qualitative difference compared to \laf , the
paramagnetic susceptibility is very similar and increases linearly with temperature. We note that
deviations from linearity as discussed for \laf , i.e. structural, magnetic and electronic
fluctuations above \ts\ in the doping regime $x\leq 0.04$ and superposition of an upturn in the
vicinity of $T_{\rm C}$ for $x\geq 0.06$, are either absent or much less pronounced in \cax .
Remarkably, the data not only show a striking qualitative similarity of $\chi_{\rm norm}$, but the
slope $\partial \chi /\partial T$ at 300\,K even quantitatively agrees in both materials as shown
in \figref{fig3}. This finding strongly suggests that the increase of $\chi_{\rm norm}$ is a
robust characteristic property of FeAs-layers in iron pnictides, with both SDW and superconducting
ground states and for 1111- as well as 122-materials.

The normal state static magnetic properties in iron pnictides and in particular the slope
$\partial \chi (300\,K)/\partial T$ do not significantly depend on the doping level $x$ or the
actual material under study, i.e. $\partial \chi$ (300\,K)$/\partial T\approx 5\cdot 10^{-7}$
erg/(G$^2$mol$_{\rm As}$K). This holds both for \laf\ where the ground state completely changes
from an afm poor metal to a non-magnetic SC and for \cax\ which exhibits an inhomogeneous
superconducting state. Neither the suppression of magnetic order and of the structural phase
transition nor the evolution of the superconducting ground state strongly affect $\partial \chi
/\partial T$ well above the ordering temperatures. The experimental fact that such a robust
feature is present in a large doping range as well as in different materials renders it very
unlikely that it depends on subtle details of the Fermi-surface and singular electronic effects.
Although we cannot completely exclude any approach critically depending on well-defined nesting
conditions, particular band filling or unique band effects, our present data do not affirm such
scenarios. This conclusion is corroborated by the absolute size of the observed susceptibility
changes. In LaFeAsO, the susceptibility changes between 150K and 500K amount to $\Delta \chi
\approx 1.9\cdot 10^{-4}$\,erg/(G$^2$mol). The electronic density of states from band structure
calculations, $N(E_{\rm F})$, imply a pure Pauli-susceptibility $\chi_{\rm P}\sim 5.5-7.5\cdot
10^{-5}$\,erg/(G$^2$mol) only.~\cite{Boeri2008,Xu2008,Haule2008,Anisimov09} Taking into account
the moderate mass enhancement of $\lesssim 2$ (Ref.~[\onlinecite{Nomura2008,Anisimov09}]), this is
too small to account for the experimentally observed $\Delta \chi$, even if changes below 150K and
above 500K, the robustness of the feature against doping and the size of the susceptibility itself
are not considered.

In contrast, we suggest that the magnetic interactions are relevant. The static susceptibility
measures the size of magnetization of the material in response to an external magnetic field.
Since, as justified above, the measured macroscopic $\chi$ in the studied iron pnictide samples is
the intrinsic spin susceptibility, the experimental data hence imply that the moment per Fe site
aligned by the constant external field becomes smaller upon cooling. Such a behavior is typical
for the evolution of local afm correlations in systems which are characterized by a spin gap.
Typical examples where spin-singlet formation yields $\partial \chi /\partial T > 0$ are spin
ladders, spin dimers, or Haldane chains.~\cite{Rice,Haldane,Misguich} Indeed, recent theories that
attempt to describe our data presented above corroborate the scenario of local afm
correlations.~\cite{Singh2009PhysicaC,Laad,Graser,Berciu2009PRB,Korotin2008,Korshunov2008,Korshunov2009PRL}
Even though contradicting approaches are used, starting from the weak or the strong coupling
limit, respectively, these studies agree that the experimentally observed increasing $\chi(T)$
implies the presence of an unusual metallic state with local afm correlations that persist up to
high temperature. Zhang et al. apply the 2D frustrated afm Heisenberg model and demonstrate that
the data in \figref{fig2} provide strong evidence for the existence of a wide afm fluctuation
window of \emph{local} magnetic moments.~\cite{Zhang2009EPL} An itinerant approach where nesting
boosts $q=(\pi,\pi)$ SDW magnetic correlations is suggested in
Refs.~[\onlinecite{Korotin2008,Korshunov2008,Korshunov2009PRL}]. In particular, Korshunov et al.
show that the linear $T$-dependence of $\chi(T)$ is universal in 2D Fermi-liquids. Remarkably,
there is a quantitative agreement between calculated and measured slopes $\partial \chi /\partial
T$ for all doping levels.~\cite{Korshunov2009PRL} While this agreement provides evidence for the
itinerant approach applied in Ref.~\onlinecite{Korshunov2009PRL}, the authors identify our
experimental data the first observation of a nonanalytic behavior of the 2D spin susceptibility.
In contrast, the data in \figref{fig2} have been taken as ample evidence for singlet pair
formation at elevated temperature in investigations of polaron formation including their dynamics
and tendency for binding.~\cite{Berciu2009PRB,Sawatzky} It is argued that preformed singlet pairs
with a binding energy of 100~meV or more would straightforwardly explain the experimental results.

We note, however, that the static susceptibility data alone do not exclude an electronic origin of
the unusual temperature dependence $\chi(T)$. In addition, a relation of the $T$-dependence of the
NMR quantities $K_s$ and $T_1^{-1}$ to a possible pseudospin gap in the normal state has been
controversially discussed in NMR works~\cite{Ahilan2008,Nakai2008,Grafe2008}. The experimental
observation of a nearly constant value $\partial \chi /\partial T$ would, however, lead to the
remarkable conclusion of a doping independent pseudogap in the complete doping range $0\leq x\leq
0.15$. Such a behavior would be completely different to the findings in the superconducting
cuprates.~\cite{Timusk}

Following the scenario of spin-singlet formation, our experimental results suggest local afm
fluctuations in the complete doping regime under study. The doping independent positive slope of
$\chi(T)$ straightforwardly implies that not only the character of these local magnetic
fluctuations but also the strength is preserved even if the ground state changes from the afm
ordered to the SC one. In this context we emphasize that \laf\ is the only pnictide system which
exhibits a homogeneous superconducting phase and the absence of any magnetic and/or structural
order in the entire normal state. Thus it is particularly suited to study intrinsic electronic and
magnetic properties avoiding complications due to inhomogeneities.


In conclusion, we have studied the static magnetic properties of \laf\ and \cax\ upon doping. Our
data confirm suppression of the structural transition and afm SDW formation and evolution of SC.
Our main result is the observation of an increasing paramagnetic susceptibility whose slope is
independent of the doping level and the material. Its intrinsic nature is confirmed by
measurements of the NMR Knight shift. Our data suggest that strong local afm correlations are
present in a broad region of the phase diagrams of iron pnictides and in particular persist in the
normal state of superconducting materials.

\begin{acknowledgments}
We thank M. Deutschmann, S. M\"uller-Litvanyi, R. M\"uller, J. Werner, K.~Leger, and S. Ga{\ss} for
technical support. Work was supported by the DFG through FOR 538 and project BE1749/12.
\end{acknowledgments}


\end{document}